\documentclass[useAMS,usenatbib]{mn2e}
\usepackage{psfig, epsf, epsfig}
\title[Globular cluster  kinematics in M31]{Origin of rotational
kinematics in the globular cluster system of M31:
A new clue to the bulge formation}

\author[K. Bekki]
{Kenji Bekki${}^1$\thanks{E-mail:
bekki@phys.unsw.edu.au}\\
       ${}^1$School of Physics, University of New South Wales,
              Sydney, NSW,  2052,  Australia\\}

\begin{document}

\date{Accepted, Received 2005 February 20; in original form }

\pagerange{\pageref{firstpage}--\pageref{lastpage}} \pubyear{2005}

\maketitle

\label{firstpage}

\begin{abstract}

We propose that   the rotational kinematics of the globular
cluster system (GCS) in M31 
can result from a  past major merger event  that could have
formed its bulge component.
We numerically investigate kinematical properties of globular
clusters (GCs) in remnants of galaxy mergers between two disks
with GCs in both  their disk and halo components. 
We  find that  the GCS 
formed during major merging
can show strongly rotational kinematics with
the maximum  rotational velocities
of $\sim 140 - 170$ km s$^{-1}$
for a certain range of orbital parameters of merging.
We also find that a rotating stellar bar, which can be morphologically
identified as a boxy bulge if seen edge-on,
can be formed in models for which the GCSs show strongly rotational kinematics.
We thus suggest that the observed rotational kinematics of
GCs with different metallicities  in M31 can be closely
associated with the ancient major merger event.
We discuss whether 
the formation of the rotating bulge/bar in M31 
can be due to  the ancient merger. 

\end{abstract}


\section{Introduction}

The structural and kinematical properties of GCSs
are considered to provide valuable information on 
the formation and evolution of their host galaxies
(e.g., Searle \& Zinn 1978; Romanowsky et al. 2009).
Recent observational studies on kinematical
properties of GCSs in galaxies and their comparison
with theoretical and numerical works have advanced 
our understanding on mass distributions of galaxies
(e.g., Pierce et al. 2006;
Romanowsky et al. 2009),  
formation of elliptical galaxies (e.g., Bekki et al. 2005), and
formation and evolution of dwarf galaxies (e.g., Beasley et al. 2009).
A growing number of observational data sets on kinematical
properties of GCSs have been now accumulated not only for
galaxies in the local group  but also 
for nearby galaxies (Brodie \& Strader 2006 for a recent
review)
so that we can discuss formation processes of their host
galaxies and their dependences
on galactic global properties
(e.g., Hubble types)   in more detail based on the observations.

One of the intriguing  properties of GCSs in galaxies
is the observed rotational kinematics of the GCS in M31 
(e.g., Huchra et al. 1982;
Perrett et al. 2002).
Recent wide-field surveys of the GCS in M31 (Lee et al. 2008) have confirmed 
that the GCS composed both of metal-poor ([Fe/H] $<-0.9$ in their
criterion) and metal-rich (Fe/H] $>-0.9$) objects has a rotational
amplitude of $\sim 190$ km s$^{-1}$.
Such  kinematics are in a striking contrast with
the low rotation of the Galaxy's GCS (e.g.,  Armandroff 1989)
with an origin 
that  remains unclear. \\

The purpose of this paper is to discuss (i) why the GCS of M31
shows such a large amount of rotation and (ii) what implications it
has for the formation and evolution of M31.
Previous  numerical works showed that 
rotational kinematics of GCSs in elliptical galaxies
can be due to past major merger events that formed the galaxies
(e.g., Hernquist \& Bolte 1993; Bekki et al. 2005).
We thus consider that the rotational  kinematics of the GCS is closely
associated with a major merger event in M31 long  ago
and thereby investigate whether galaxy merging can reproduce well
the observed kinematics of the GCS.
We also discuss whether the observed rotating bulge/bar in the M31
(e.g., Beaton et al. 2007)
can be due to the major merger event responsible for the formation
of the GCS with rotational kinematics.
\begin{table*}
\centering
\begin{minipage}{185mm}
\caption{The values of model parameters and a brief summary of the
results}
\begin{tabular}{ccccccccc}
model &
{$M_{\rm dm}/M_{\rm d}$
\footnote{The mass ratio of dark matter halo to stellar disk 
in a galaxy.}}&
{$m_{\rm 2}$
\footnote{The mass ratio  of two disks 
in a galaxy merger.}}&
{$r_{\rm p}$ ($\times R_{\rm d}$)
\footnote{The pericenter distance of 
a merger in units of the disk size $R_{\rm d}$. }}&
{$e_{\rm p}$
\footnote{The orbital eccentricity of
a merger.}}&
{orbital
\footnote{``PP'' and ``PR'' represent prograde-prograde and
prograde-retrograde merging, respectively. }} &
{$V_{\rm max,dgc}$ (km s$^{-1}$)
\footnote{The maximum rotational velocity of DGCs  (GCs initially
in disks)
in a merger remnant. }} &
{$V_{\rm max,hgc}$ (km s$^{-1}$)
\footnote{The maximum rotational velocity of HGCs 
(GCs initially in halos)
in a merger remnant. }} &
comments \\
1 & 9 & 1.0 & 2.0  & 0.72  & PP & 136 & 123 & the standard model \\
2 & 9 & 1.0 & 2.0  & 0.72  & PR & 114 & 133 &  \\
3 & 9 & 1.0 & 1.0  & 0.85  & PP & 95 & 92 &  \\
4 & 9 & 0.5 & 2.0  & 0.72  & PP & 107 & 38 &  unequal-mass merger\\
5 & 19 & 1.0 & 2.0  & 0.72  & PP & 174 & 91 &  \\
\end{tabular}
\end{minipage}
\end{table*}

\section{The model}

Since the numerical methods and techniques we employ for modeling
dynamical evolution of  mergers 
between two disks with GCs have already been detailed
elsewhere (Bekki et al. 2005; Bekki \& Forbes 2006), we give only  a brief review here.
The progenitor disk galaxies that take part in a merger are given
a dark halo,  a thin exponential disk,
and GCs initially in disks (referred to as ``DGCs'' from now on)
and in halos (``HGCs'').
The total mass and size of an exponential
disk with no bulge 
are $M_{\rm d}$ and $R_{\rm d}$,
respectively.

We consider 
that the total stellar mass of a merger remnant
should be similar to the total mass ($M_{\rm b}$) of the present bulge of M31
($M_{\rm b}=3-4 \times 10^{10} {\rm M}_{\odot}$;
e.g., Seigar et al. 2008; Geehan et al. 2006). 
We thus adopt 
$M_{\rm d}=2 \times 10^{10} {\rm M}_{\odot}$
as a  reasonable value in the present study.
We adopt the density distribution of the NFW
halo (Navarro, Frenk \& White 1996) with the mass of $M_{\rm dm}$
and a concentration parameter determined by  $M_{\rm dm}$ 
according to the formula derived by  recent CDM simulations (e.g., Neto
et al 2007). The stellar disk has the scale length ($R_0$)
of $0.2R_{\rm d}$. 

\begin{figure}
\psfig{file=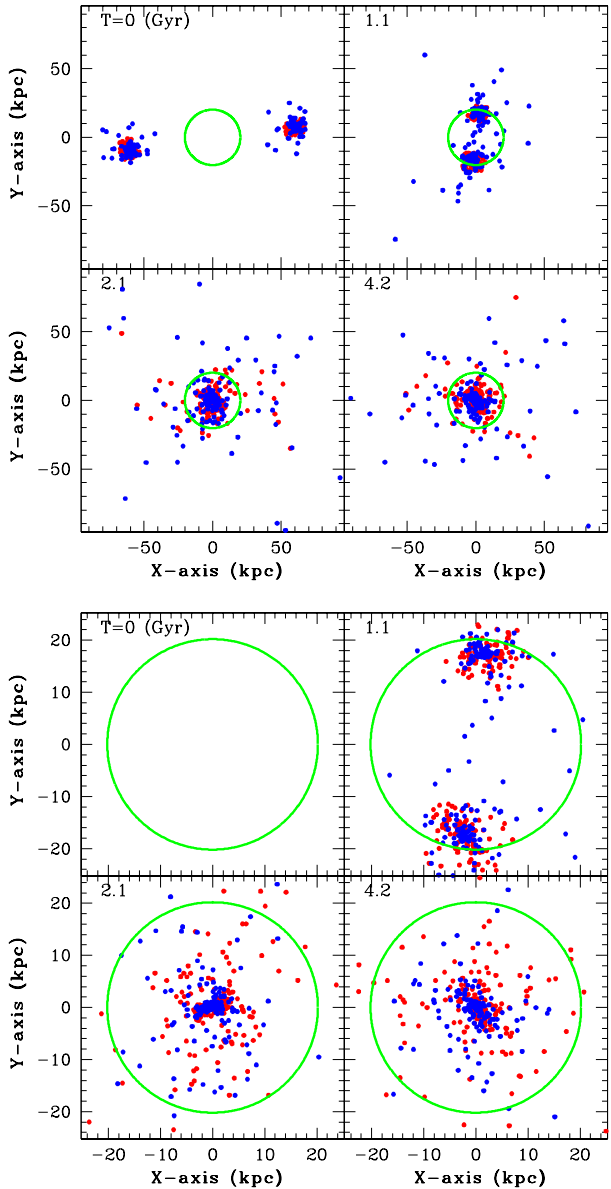,width=8.0cm}
\caption{
Distributions of disk GCs (DGC, red) and halo ones (HGC, blue) of
a galaxy merger 
projected onto the $x$-$y$ plane (i.e., orbital plane of the merger)
for larger (upper) and smaller (lower) scales of view 
in the standard model. The time $T$ shown in the upper left
corner of each panel indicates the time that elapsed since
the simulation starts. Green circles indicate 20 kpc from the
mass center of the merger. 
}
\label{Figure. 1}
\end{figure}

The present bulge-less disk model would be reasonable for 
less massive disk galaxies with no/little bulges
like the Large Magellanic Cloud and Magellanic-type galaxies
with  total masses  of $\sim 10^{10} {\rm M}_{\odot}$,
but it would not be so realistic  for luminous disk galaxies like
the merger progenitor disks of M31.
We however conjecture since the bulge masses are quite small in comparison
with total masses of galaxies (inclusive of dark matter halos),
final kinematics of GCSs in merger remnants would not depend strongly
on  whether or not we include bulges in initial disks.
Thus,  as long as we mainly discuss global kinematical properties
of GCSs in merger remnants,  
the present model can be regarded as an appropriate  one.

Previous observations revealed that  less luminous
disk galaxies, which are considered to form the bulge component
of M31 with rotational kinematics of the GCS
in the present study,  have disky distributions with rotational
kinematics in their GCs (e.g., Freeman et al. 1983; Olsen et al. 2004).
We thus consider the presence of DGCs in merger progenitor  disks
in the present study: DGCs were  not considered in our previous
works (e.g., Bekki et al. 2003a; 2005) and such DGCs with
strongly rotational kinematics  
are not observed in the metal-rich GCs  of the Galaxy.
The DGCs have the same exponential distribution
and rotational kinematics  as field stars
in the stellar disk of their host galaxy.
The initial rotational amplitude  of DGCs  is $\sim 170$ km s$^{-1}$
for $M_{\rm d}=2 \times 10^{10} {\rm M}_{\odot}$ and
$M_{\rm dm}/M_{\rm d}=9$.

The Galactic HGCs and the stellar halo
have similar radial density profiles of ${\rho}(r)$ $\propto$ $r^{-3.5}$
(van den Bergh 2000).
We therefore assume that the HGCs in our galaxies have a power-law profile
with an  exponent of $-3.5$ and a  half-number radius of $1.4 R_{0}$
(which is $\sim 5$  kpc for the Galaxy and thus consistent with
observations). The HGCs are assumed to have isotropic velocity
dispersions determined by the mass distribution of the galaxy.
Total numbers of DGCs and HGCs in a galaxy
are set to be 100 and 100, respectively.

The mass ratio of the two disks  ($m_2$) in  a merger is
assumed to be a free parameter.
In all of the simulations of pair mergers, the orbit of the two disks is
set to be
initially in the $xy$ plane and the distance between
the center of mass of the two disks
is  assumed  to be $12R_{\rm d}$.
The pericenter distance ($r_{\rm p}$) and the eccentricity ($e_{\rm p}$)
in a pair merger are assumed  to be free parameters that control
orbital energy and angular momentum of the merger.
The spin of each galaxy in a merger
is specified by two angles $\theta_{i}$ and
$\phi_{i}$, where suffix  $i$ is used to identify each galaxy.
$\theta_{i}$ is the angle between the $z$ axis and the vector of
the angular momentum of a disk.
$\phi_{i}$ is the azimuthal angle measured from the $x$ axis to
the projection of the angular momentum vector of a disk onto the $xy$
plane.

We mainly show the results of ``the standard model'' with 
$M_{\rm d}=2 \times 10^{10} {\rm M}_{\odot}$,
$M_{\rm dm}/M_{\rm d}=9$, $m_{\rm 2}=1$,
$r_{\rm p}=2R_{\rm d}$, $e_{\rm p}=0.72$,
$\theta_{1}=30^{\circ}$,
$\theta_{2}=45^{\circ}$,
$\phi_{1}=45^{\circ}$, and
$\phi_{2}=120^{\circ}$.
We describe the results of this standard model
with an orbital configuration similar
to ``prograde-prograde'' merging (in which the orbital spin axis of 
the merger is parallel to intrinsic spin axes of the two disks),
mainly because the final GCS can clearly show strongly rotational kinematics
as observed in M31.
We also show the results of the model with a ``prograde-retrograde''
orbital configuration in which $\theta_{2}=225^{\circ}$ and
other parameters are exactly the same as those in the standard model.

In order to discuss the origin of the central bar/bulge of M31,
we investigate in which models
the merger remnants show rotating stellar bars.
Although we investigate 34 models, we show the results of five
representative models which either show  clearly rotating bars
or have no such rotating components in merger remnants: comparison
between these models enables us to understand better the role
of an ancient merger event in the formation of the bulge component
in M31. 

The model parameters and some brief
results are given in the Table 1.
The total particle number for a major  merger in a simulation is $10^6$ 
and the simulation is  carried out on  
the latest version of GRAPE (GRavity PipE, GRAPE-7) which is the
special-purpose
computer for gravitational dynamics (Sugimoto et al. 1990).
In estimating the GCS kinematics, the merger remnant is viewed near
to edge-on, and binned major-axis profiles are constructed. In order
to have enough objects in each bin, GCs at all minor axis distances
are
included, which also approximately replicates the comparison
observations.
Formation of metal-rich GCs through dissipative gas dynamics
of galaxy merging  (e.g., Bekki et al. 2002) and 
adiabatic contraction of the GCS in M31 due to later massive growth
of the disk component could increase appreciably
the rotational amplitude $V$ of the GCS.
The quantitative estimation of this effect is not done here  and will
be done in our future studies.
It should be stressed that this paper is meant to be schematic
rather than reproducing the observed galaxy in a fully self-consistent
manner.

\begin{figure}
\psfig{file=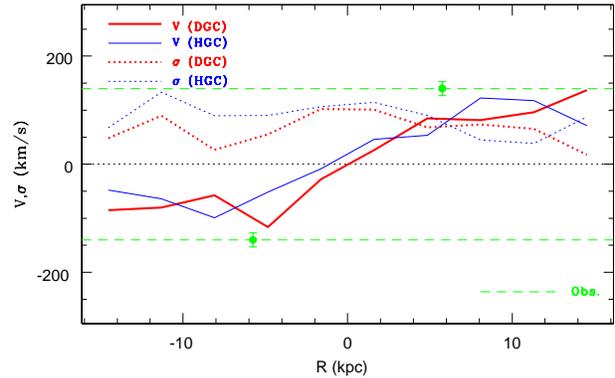,width=8.0cm}
\caption{
Radial profiles of line-of-sight velocities ($V$, solid) and
velocity dispersions ($\sigma$, dotted) for DGCs (red) and HGCs (blue)
for the merger remnant at $T=4.2$ Gyr
in the standard model.
The dotted horizontal black line represents $V$ ($\sigma$) = 0 km s$^{-1}$.
The observed $v_{\rm rot} \sim 140$ km s$^{-1}$  for
the full sample of GCs in M31 (Perrett et al.
2002) is shown by a dashed green line for comparison: Lee et al (2008)
show significantly larger $v_{\rm rot}$ ($\sim 188$ km s$^{-1}$) 
of the GCS for $|Y| \ge$ 0 kpc.
The locations of the filled circles indicate where the observed velocity
profile becomes flat and the error bars are observational ones
($\pm 13$ km s$^{-1}$).
}
\label{Figure. 2}
\end{figure}

\begin{figure}
\psfig{file=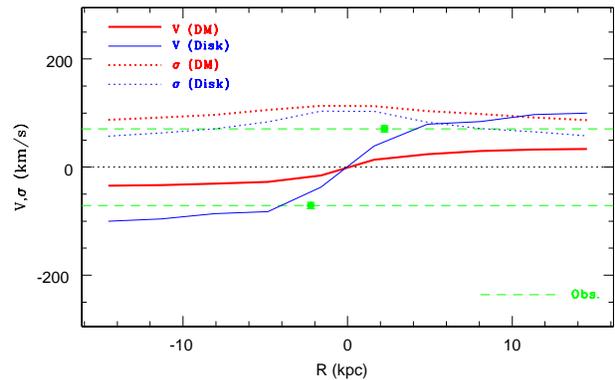,width=8.0cm}
\caption{
The same as Fig. 2 but for 
for dark matter  (red) and 
disk stars (blue).
The bulge observational results are from McElroy (1983) and
the locations of the filled circles indicate the 
observed outermost
region ($R=9.76^{'}$) at the position angle of 45$^{\circ}$.
}
\label{Figure. 3}
\end{figure}

\begin{figure}
\psfig{file=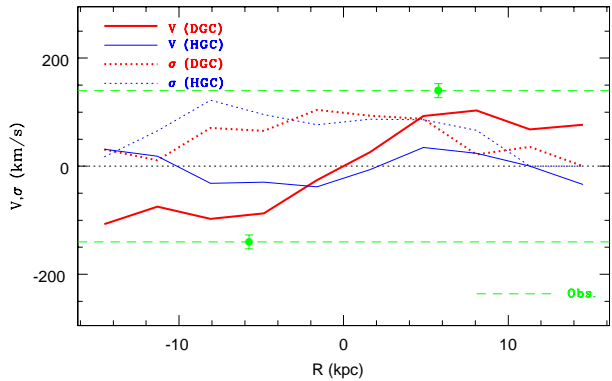,width=8.0cm}
\caption{
The same as Fig. 2 but for the model with $m_{\rm 2}=0.5$.
}
\label{Figure. 5}
\end{figure}

\begin{figure}
\psfig{file=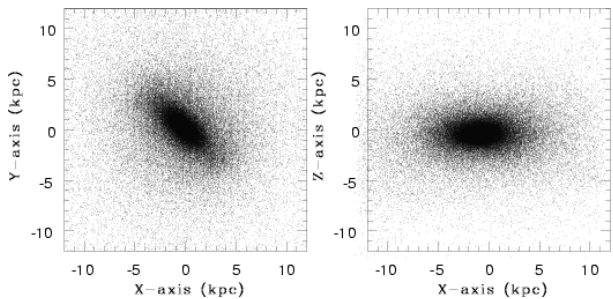,width=8.0cm}
\caption{
Distributions of stars projected onto the $x$-$y$ (left) 
and $x$-$z$ (right) planes
at $T=4.2$ Gyr
in the standard model.
Only one for every ten particles is shown for clarity.
}
\label{Figure. 5}
\end{figure}

\section{Results}

Fig. 1 shows that both DGCs and HGCs can be 
spatially mixed to form a new GCS 
during violent dynamical relation that results in transformation
from two disks into one spheroidal galaxy. 
Owing to angular momentum redistribution during major merging,
not only HGCs but also DGCs can be 
transferred to the outer halo region and thus seen there 
in the merger remnant. The half-number radii for the DGCs and 
the HGCs are both $\sim  5.0$  kpc in the merger
remnant, which is consistent with the observations
by Battistini et al. (1993).
As the stellar remnant shows a rotating bar (discussed later in \S 4),
DGCs and HGCs also show barred structures, appreciably flattened shapes
(qualitatively consistent with observations),
and figure rotation.

Fig. 2 clearly shows global rotation both in DGCs and HGCs, though
the radial profiles of $V$ and $\sigma$ do  not smoothly change
owing to the small numbers of GCs at each bin.
The maximum $V$ ($V_{\rm max}$) and $\sigma$ (${\sigma}_{\rm max}$)
for DGCs (HGCs) within the central 16 kpc are  
136 km s$^{-1}$ (123 km s$^{-1}$)
and 102 km s$^{-1}$ (106 km s$^{-1}$), respectively.
Therefore $V_{\rm max}/{\sigma}_{\rm max}$ for DGCs and HGCs
are 1.3 and 1.2, respectively, which means that the final GCS
has strongly rotational kinematics.
The velocity profile becomes flat ($V\sim 100$ km s$^{-1}$)
at $R \sim 4$ kpc.
The simulated two-dimensional line-of-sight velocity
map  of the GCS shows clearly global rotation.

We here compare the simulated rotational kinematics with the observed
one in M31
(e.g., 
$V =138 \pm 13$ km s$^{-1}$ in Perrett et al. 2002).
Observational error bars in $V$
are not  so small
and $V$ ranges from 98 km s$^{-1}$ to
188 km s$^{-1}$ for $0$kpc $\le |Y| \le$ 5kpc
(Lee et al.  2008),
where $|Y|$ is the projected vertical distance from the
M31 disk plane.
The simulated rational amplitude of the GCS is slightly
smaller than the observed one by  Perrett et al. (2002): It should
be noted that the observational results depend on the details
of how the data are binned and fitted.

We consider that the best model needs to 
reproduce the above $V$  of $\sim 140$ km s$^{-1}$
Fig. 3 shows that even the dark matter halo
of the merger remnant can have a small amount of rotation
($V \sim 34$ km s$^{-1}$) owing to the redistribution
of angular momentum (i.e., conversion of orbital  angular momentum
of merging two galaxies 
into internal one of the merger remnant). 
This result suggests that if the two galaxies have hot gaseous halos,
then the remnant is highly likely to have a slowly rotating gaseous
halo. 
Fig. 3 also shows that the stellar component of the merger remnant
has a significant amount of rotation ($V_{\rm max} \sim 100$ km
s$^{-1}$), which means that the spheroid is  rapidly rotating
(i.e., rotating bulge is formed  from major merging).

The reason for the smaller $V$ of the dark matter halo in comparison
with the GCS is due largely to the difference in the initial spatial
distribution between the halo and the GCS, demonstrating
that GC rotation would  not be used to infer dark matter halo rotation.
The inner part of the merger can more strongly spin-up
during and after major merging owing to (i)  the prograde-prograde orbital
configuration of the merging  and (ii) the development of the rotating bar. 
The distribution
of the GCS in the merger
progenitor disk is by a factor of 4 more compact that that of
the halo so that the GCS can more strongly spin-up: most of the individual
GCs can obtain a larger
amount of intrinsic spin  angular momentum with respect to the center of
the merger remnant.

Fig. 4 shows that strongly rotational kinematics of GCs can be seen
only in DGCs in the model with $m_{2}=0.5$ (Model 4): $V_{\rm max}$
is 107 km s$^{-1}$ for DGCs and 38 km s$^{-1}$ for HGCs.
The higher and lower $V_{\rm max}$
in DGCs and HGCs, respectively, are confirmed in
other unequal-mass merger models (e.g., $m_2=0.3$).
It should be noted here that the model 1 shows
only slightly  higher $V_{\rm max}$ in DGCs, which 
is due to different kinematics between DGCs and HGCs
(i.e., {\it initial} global  rotation only in DGCs). 
These results therefore mean  
that if metal-poor and metal-rich  GCs originate from 
HGCs and DGCs, respectively,  then kinematical properties of GCs
in the remnants of galaxy merging with smaller $m_{2}$ (or
unequal-mass merging)
can be significantly different between
metal-poor and metal-rich ones.  
These results furthermore imply that the observed rotational
kinematics both for metal-poor and metal-rich GCs in M31
can give some constraints on the mass-ratio ($m_2$) of
two disks in galaxy merging that could have occurred in M31.

The model 1 does not explain well the observed total halo
mass:
later numerous accretion events of dwarfs with few GCs and stars
are required to increase significantly  the total mass after
the  merging.
The models with larger $M_{\rm dm}/M_{\rm d}$ can show higher
$V_{\rm max}$ in DGCs owing to the initially higher circular velocities
of the stellar disks. For example, the model with $M_{\rm dm}/M_{\rm d}=19$
(Model 5)
shows $V_{\rm max}=174$ km s$^{-1}$ and ${\sigma}_{\rm max}=134$ km
s$^{-1}$: it should be stressed here that the total mass of the remnant
is $8\times 10^{11} {\rm M}_{\odot}$
thus can be 
more consistent with observations 
than Model 1 in terms of the total mass of M31.
A possible reason for
the lower $V_{\rm max}$ (91 km s$^{-1}$) for HGCs in the model 5 
is that more strongly self-gravitating
stellar disks can also play a role in increasing
global rotation of HGCs: Such a role  is weaker in the model 5 
in which the disk is much more weakly self-gravitating.

Models with different orbital configurations can show stronger
rotational kinematics in DGCs and HGCs, if larger $r_{\rm p}$ are
adopted. For example,
the model with a prograde-retrograde orbital configuration (Model 2)
shows $V_{\rm max}=114$ km s$^{-1}$ and ${\sigma}_{\rm max}=102$ km 
s$^{-1}$ but does not show a bar.
The models with smaller $r_{\rm p}$ (e.g., Model 3)
show smaller $V_{\rm max}$ in DGCs, which implies that the observed
$V_{\rm max}$ can give some constraints on the  orbital parameters
for galaxy merging that occurred in  M31.
Thus only the models with larger $r_{\rm p}$ can better reproduce
the observed kinematical properties of the M31 GCS.

\section{Discussion and conclusions}

Fig. 5 shows that the stellar remnant looks like a bar if it is  viewed 
from face-on in the standard model
of the present study: the bar is confirmed to have figure rotation.
The stellar distribution viewed from edge-on  
appears to have a flattened  spherical body, which can be identified as
a bulge.
The present study thus implies that M31's observed bulge/bar
(e.g., Beaton et al. 2007)
 can be formed from
an ancient major merger event. 
It should be noted here that 
dissipative gas dynamics which can determine the final morphological
properties of merger remnants (e.g., boxy or disky shapes)
is not included in the present study.

If the inner bar/bulge of M31 was really formed from  ancient major
merging before its disk formation,
then the later development of the stellar and gaseous disk
may well be significantly influenced by dynamical action of the already
formed bar.
Also, later slow gas accretion and the resultant disk formation in M31
could change  structural and kinematics properties of the already
developed GCS
to some extent: it would be possible that adiabatic compression
of the GCS by later gradual development of the disk can enhance
the rotational amplitude of the GCS.
It is our future study to numerically investigate
disk formation and evolution
of M31  under the presence of the already formed bar.

The present work suggests that   the hot diffuse halo gas recently detected by
{\it Chandra} (Li \& Wang 2007) can have a significant amount
of rotation resulting from angular momentum redistribution of
the possible major merger.
Furthermore, the observed extensive HI cloud population of M31
(e.g., Thilker et al. 2004; Westmeier et al. 2007)
can also have rotational kinematics if they originate from
stripped HI gas from the merger progenitor disks.
Given that
major merging can form 
very extended stellar halos (Bekki et al. 2003b;
Bekki \& Peng 2006),
the observed extended stellar halo in M31 (e.g., Ibata et al. 2007)
would have fossil
information in the possible ancient major merger event.

If M31 was formed from multiple and sequential
merging of dwarf satellites 
from random directions, it seems to be highly unlikely that
the final GCS 
has  a large amount
of global rotation. 
Therefore, 
if the rotational kinematics of the GCS derived from
previous observations 
is further confirmed by ongoing observational studies,
it then 
suggests that a dramatic physical process is responsible 
for the observed {\it globally organized motion}  of GCs.
Our simulations imply  that an ancient major merger of
two disks with GCs with different metallicities
could be  responsible for the enigmatic kinematics of  the GCS in M31,
as well as the inner bar or bulge.

\section{Acknowledgment}
I am   grateful to the anonymous referee for valuable comments,
which contribute to improve the present paper.


\begin{thebibliography}{}

\bibitem[]{}
Armandroff, T. E., 
1989, AJ, 97, 375

\bibitem[]{}
Battistini, P. L.,  Bonoli, F.,  Casavecchia, M., Ciotti, L.,  Federici,
L.,  Fusi-Pecci, F.,
1993, A\&A, 272, 77

\bibitem[]{}
Beasley, M. A., Cenarro, A. J.,  Strader, J.,  Brodie, J. P.,
2009, AJ, 137, 5146

\bibitem[]{}
Beaton, R. L., et al. 
2007, ApJ, 658, 91

\bibitem[]{}
Bekki, K., Forbes, D. A.,  Beasley, M. A., \&  Couch, W. J.
2002, MNRAS, 344, 1334

\bibitem[]{}
Bekki, K., Forbes, D. A.,  Beasley, M. A.,   Couch, W. J.
2003a, MNRAS, 344, 1334

\bibitem{} Bekki, K., Harris, W. E.,   Harris, G. L. H.,
2003b, MNRAS,
338, 587

\bibitem[]{}
Bekki, K.,  Beasley, M. A., Brodie, J. P., \& Forbes, D. A.
2005,  MNRAS, 363, 1211


\bibitem[]{}
Bekki, K.,  Forbes, D. A.,
2006, A\&A, 445, 485


\bibitem[]{}
Bekki, K., Peng, E. W., 2006, MNRAS, 370, 1737




\bibitem[]{}
Brodie, J. P., Strader, J., 2006,  ARA\&A, 44, 193






\bibitem[]{}
Freeman, K. C., Illingworth, G., \& Oemler, A., Jr.,
1983, ApJ, 272, 488

\bibitem[]{}
Geehan, J. J.,  Fardal, M. A., Babul, A., Guhathakurta, P.,
2006, MNRAS, 366, 996


\bibitem[]{}
Hernquist, L.,  Bolte, M., 
1993, 
in The globular clusters-galaxy connection, 
ASP, v48,  p788
edited by Graeme H. Smith, and Jean P. Brodie, 

\bibitem[]{}
Huchra, J.,  Stauffer, J., van Speybroeck, L.,
1982, ApJ, 259, L57

\bibitem[]{}
Ibata, R., Martin, N. F., Irwin, M., Chapman, S.,  Ferguson, A. M. N.,
Lewis, G. F., McConnachie, A. W., 2007, ApJ, 671, 1591

\bibitem[]{}
Lee, M. G., Hwang, H. S.,  Kim, S. C.,  Park, H.  S.,
Geisler, D.,  Sarajedini, A.,  Harris, W. E.,
2008, ApJ, 674, 886

\bibitem[]{}
Li, Z.,  Wang, Q. D.,
2007, ApJ, 668, L39

\bibitem[]{}
McElroy, D. B., 
1983, ApJ, 270, 485

\bibitem[]{}
Navarro, J. F., Frenk, C. S.,  White, S. D. M.,
1996, ApJ, 462, 563 (NFW)

\bibitem[]{}
Neto, A. F., et al. 2007
MNRAS, 381, 1450

\bibitem[]{}
Olsen, K. A. G., Miller, B. W., Suntzeff, N. B.; Schommer,
Robert A.,  Bright, J.,
2004, AJ, 127, 2674

\bibitem[]{}
Perrett, K. M., et al.,
2002, AJ, 123, 2490

\bibitem[]{}
Pierce, M. et al. 
2006, MNRAS, 368, 325

\bibitem[]{}
Romanowsky, A. J., Strader, J., Spitler, L. R., Johnson, R.,
Brodie, J. P., Forbes, D. A.,  Ponman, T.,
2009, AJ, 137, 4956


\bibitem[]{} 
Searle,~L.  Zinn,~R. 1978, ApJ, 225, 357 

\bibitem[]{}
Seigar, M., S., Barth, A. J., Bullock, J. S.,
2008, MNRAS, 389, 1911


\bibitem[]{}
Sugimoto, D., Chikada, Y., Makino, J., Ito, T., Ebisuzaki, T.,
Umemura, M., 1990, Nat, 345, 33

 \bibitem[]{}
Thilker, D. A., Braun, R.,  Walterbos, R. A. M., Corbelli,
E.,  Lockman, F. J., Murphy, E.,  Maddalena, R.,
2004, ApJL, 601, 39

\bibitem[]{}
van den Bergh, S.,
2000,
The Galaxies of the Local Group, Cambridge: Cambridge Univ. Press.

\bibitem[]{}
Westmeier, T.,  Braun, R.,  Br\"uns, C.,  Kerp, J.,  Thilker, D. A.,
2004, NewAR, 51, 108





\end{thebibliography}
\end{document}